\RequirePackage[table]{xcolor}
\documentclass{PoS}
\usepackage{setspace}
\usepackage{slashed}
\usepackage{verbatim}    
\usepackage{graphicx} 
\usepackage{wrapfig}
\usepackage{dsfont}
\usepackage{epsfig}
\usepackage{epstopdf}
\usepackage{amsmath}
\usepackage{amsfonts,amssymb}
\usepackage{cleveref}
\usepackage{xcolor}

\newcommand{\be}{\begin{equation}}
\newcommand{\ee}{\end{equation}}
\newcommand{\bea}{\begin{eqnarray}} 
\newcommand{\eea}{\end{eqnarray}}
\newcommand{\MSbar}{{\overline{\rm MS}}}

\newcommand{\la}{\lambda}

\title{Supersymmetric QCD on the lattice: Fine-tuning and counterterms for the Yukawa and quartic couplings}
\ShortTitle{Yukawa and Quartic Couplings in Supersymmetric QCD}

    \author{M.~Costa$^{a,\, b, \, c}$, \speaker{H.~Herodotou}$^{ \,a}$, H.~Panagopoulos$^{\,a}$\\
	\llap{}$^a$Department of Physics, University of Cyprus, Nicosia, CY-1678, Cyprus\\
	$^b$Department of Mechanical Engineering and Material Science and Engineering, Cyprus University of Technology,  Limassol, CY-3036, Cyprus \\
    $^c$Rinnoco Ltd, Limassol, CY-3047, Cyprus \\
	{\rm E-mail}:  \email{kosta.marios@ucy.ac.cy}, \email{herodotos.herodotou@ucy.ac.cy}, \email{panagopoulos.haris@ucy.ac.cy}}
	
\abstract{In this study, we determine the fine-tuning of parameters in $\mathcal{N} = 1$ Supersymmetric QCD, discretized on a Euclidean lattice. Our focus is on the renormalization of Yukawa (gluino-quark-squark interactions) and quartic (four-squark interactions) couplings. Given that SUSY is broken on the lattice, non-supersymmetric counterterms must be added to the discretized Lagrangian, with coefficients which must be appropriately fine-tuned in order to recover SUSY in the continuum limit. To deduce the renormalization factors and the coefficients of the counterterms, we compute the relevant three-point (for the Yukawa couplings) and four-point Green's functions (for the quartic couplings) perturbatively to one-loop and to the lowest order in lattice spacing. Both dimensional and lattice regularizations are used to implement the Modified Minimal Subtraction scheme. Our lattice formulation employs the Wilson discretization for gluino and quark fields, the Wilson gauge action for gluons, and naive discretization for squark fields. One main difficulty in this study lies in the fact that different components of squark fields mix among themselves at the quantum level. Consequently, for an appropriate fine-tuning of the couplings, these mixings must be taken into account in the renormalization conditions. All Green's functions and renormalization factors exhibit an explicit dependence on the number of colors, $N_c$, the number of flavors, $N_f$, and the gauge parameter, $\alpha$, which are left unspecified. This work follows previous investigations on SQCD and finalizes the one-loop fine-tuning of the SQCD action on the lattice, paving the way for numerical simulations of SQCD.
\begin{center}
\includegraphics[scale=0.2]{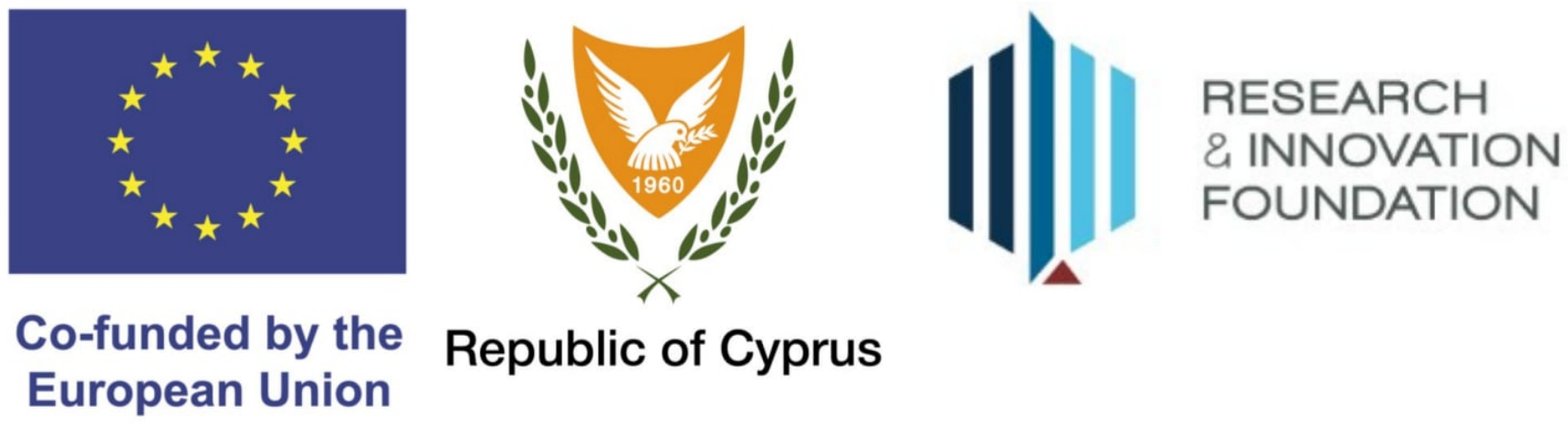}
\end{center}
}

\FullConference{%
The 41st International Symposium on Lattice Field Theory,\\
28th July-3rd August, 2024,\\
University of Liverpool, United Kingdom
}

\begin{document}
 
	\maketitle

\section{Introduction}
\setcounter{page}{2}
In recent decades, supersymmetry (SUSY) has emerged as a promising candidate for adressing several unresolved issues in the Standard Model (SM), such as understanding dark matter and the unification of electromagnetic, weak, and strong forces as proposed by the Grand Unified Theory (GUT). Researchers have been actively investigating supersymmetric theories involving strongly interacting particles to address these challenges, with significant efforts focused on studying SUSY phase transitions and SUSY breaking mechanisms through numerical lattice simulations. Lately, there has been increasing interest in exploring supersymmetric QCD (SQCD). However, a major challenge is the need for fine-tuning the bare parameters of the theory's Lagrangian. This fine-tuning is crucial to mitigate unwanted effects and ensure that the numerical results are meaningful.

This project continues previous work on SQCD and completes the one-loop fine-tuning of the SQCD action on the lattice, setting the stage for future numerical simulations of the theory. Specifically, in Refs.~\cite{Costa:2017rht} and \cite{Costa:2018mvb}, the first lattice perturbative calculations for SQCD were performed. In Refs.~\cite{Costa:2017rht} and \cite{Costa:2018mvb} we calculate the renormalization of all the parameters and fields in the supersymmetric Lagrangian using Wilson gluons and fermions, apart from the Yukawa and quartic couplings \cite{Herodotou:2022xhz, Costa:2023lna, Costa:2023cqv, Costa:2024tyz}. Additionally, the mixing of some composite operators under renormalization was explored. Our current research builds on these results to further advance the study of SQCD.

In this study we focus on the lattice renormalization of both Yukawa and quartic couplings. Our investigation employs the Wilson gauge action for gluon fields, the Wilson fermion action for quark and gluino fields, and a naïve discretization for squark fields. After outlining the key elements of our computational setup in Section \ref{comsetUP}, we examine the renormalization of Yukawa couplings in Section \ref{couplingY} within both dimensional and lattice regularization frameworks. We use the $\MSbar$ renormalization scheme to compute the renormalization factors at one-loop order. Additionally, we present our findings on quartic coupling renormalization in Section \ref{couplingQ}. We conclude with a brief summary of our work and discuss future research plans in Section \ref{summary}.
 
	\section{Formulation and Computational Setup}
	\label{comsetUP}
   
In this section we present the Euclidean SQCD action ${\cal S}^{L}_{\rm SQCD}$ on the lattice in the Wess-Zumino (WZ) gauge \cite{Costa:2017rht, Costa:2018mvb, Curci:1987, Schaich:2014, Giedt:2014, Bergner:2016} using Wilson fermions:
\begin{small}
\bea
{\cal S}^{L}_{\rm SQCD} & = & a^4 \sum_x \Big[ \frac{N_c}{g^2} \sum_{\mu,\,\nu}\left(1-\frac{1}{N_c}\, {\rm Tr} U_{\mu\,\nu} \right ) + \sum_{\mu} {\rm Tr} \left(\bar \lambda  \gamma_\mu {\cal{D}}_\mu\lambda  \right ) - a \frac{r}{2} {\rm Tr}\left(\bar \lambda   {\cal{D}}^2 \lambda  \right) \nonumber \\ 
&+&\sum_{\mu}\left( {\cal{D}}_\mu A_+^{\dagger}{\cal{D}}_\mu A_+ + {\cal{D}}_\mu A_- {\cal{D}}_\mu A_-^{\dagger}+ \bar \psi  \gamma_\mu {\cal{D}}_\mu \psi  \right) - a \frac{r}{2} \bar \psi   {\cal{D}}^2 \psi  \nonumber \\
&+&i \sqrt2 g \big( A^{\dagger}_+ \bar{\lambda}^{\alpha}  T^{\alpha} P_+ \psi   -  \bar{\psi}  P_- \lambda^{\alpha}   T^{\alpha} A_+ +  A_- \bar{\lambda}^{\alpha}  T^{\alpha} P_- \psi   -  \bar{\psi}  P_+ \lambda^{\alpha}   T^{\alpha} A_-^{\dagger}\big)\nonumber\\  
&+& \frac{1}{2} g^2 (A^{\dagger}_+ T^{\alpha} A_+ -  A_- T^{\alpha} A^{\dagger}_-)^2 - m ( \bar \psi  \psi  - m A^{\dagger}_+ A_+  - m A_- A^{\dagger}_-)
\Big] \,.
\label{susylagrLattice}
\eea
\end{small}In our calculations, we adopt standard discretization, where quarks ($\psi$), squarks ($A_\pm$), and gluinos ($\lambda$) are defined on the lattice points, while gluons ($u_\mu$) are defined on the links between adjacent points: $U_\mu (x) = \text{exp}[i g a T^{\alpha} u_\mu^\alpha (x+a\hat{\mu}/2)]$, with $\alpha$ serving as the color index in the adjoint representation of the gauge group. In Eq.~(\ref{susylagrLattice}), $P_\pm$ are projectors: $P_\pm= (1 \pm \,\gamma_5)/2$, $a$ is the lattice spacing, $U_{\mu \nu}(x) =U_\mu(x)U_\nu(x+a\hat\mu)U^\dagger_\mu(x+a\hat\nu)U_\nu^\dagger(x)$, $T^{\alpha}$ are the $SU(N_c)$ generators in the fundamental representation, $m$ is the mass of the matter fields (which may be flavor-dependent), $\cal{D}$ is the standard covariant derivative in the fundamental/adjoint representation \cite{Costa:2017rht}, $r$ is the Wilson parameter, $N_c$ is the number of colors. A summation over flavors is understood in the last three lines of Eq.~(\ref{susylagrLattice}). 


There are symmetries of the continuum theory, such as parity ($\cal{P}$) and charge conjugation ($\cal{C}$), which are preserved exactly in the lattice formulation.  Additionally, the classical action exhibits further symmetries: $\cal{R}$, which represents a $U(1)_R$ rotation of the quark and gluino fields, and $\cal{\chi}$, a $U(1)_A$ axial rotation of the squark doublets ($A_+$, $A_-$) and quark fields. However, both $\cal{R}$ and $\cal{\chi}$ symmetries are broken on the lattice due to the presence of Wilson terms. The transformations of the fields under these symmetries are detailed in Refs.~\cite{Herodotou:2022xhz, Costa:2023cqv}.

	\section{Renormalization of the Yukawa Couplings}
	\label{couplingY}

To explore the renormalization of Yukawa couplings, we analyze how dimension-4 operators, which are gauge-invariant, flavor singlets, and involve a gluino, quark, and squark field, transform under the symmetries $\cal{P}$ and $\cal{C}$. From this analysis, we determine that there are two linear combinations of Yukawa-type operators, which remain invariant under both $\cal{P}$ and $\cal{C}$ symmetries~\cite{Giedt:2009}:
\bea
\label{chiInv}
&&Y_1 \equiv A^{\dagger}_+ \bar{\lambda} P_+ \psi   -  \bar{\psi}  P_- \lambda A_+ +  A_- \bar{\lambda} P_- \psi   -  \bar{\psi}  P_+ \lambda A_-^{\dagger} \, ,\\
&&Y_2 \equiv A^{\dagger}_+ \bar{\lambda} P_- \psi   -  \bar{\psi}  P_+ \lambda A_+ +  A_- \bar{\lambda} P_+ \psi   -  \bar{\psi}  P_- \lambda A_-^{\dagger} \, .
\label{chiNonInv}
\eea

Thus, each term within the combinations, $Y_1$ and $Y_2$, is associated with a single Yukawa coupling, denoted as $g_{Y_1}$ and $g_{Y_2}$, respectively. Notably, the first combination corresponds to the third line in Eq.~(\ref{susylagrLattice}). However, at the quantum level, the second, mirror combination can also arise, with its own Yukawa coupling, $g_{Y_2}$. In the classical continuum limit, $g_{Y_1}$ coincides with the gauge coupling $g$, while $g_{Y_2}$ vanishes.




Utilizing the renormalization factors for the fields and the gauge coupling in $DR$ \cite{Costa:2017rht}, we obtain the result for the renormalization factor $Z_{Y_1}^{DR, \MSbar}$ of the Yukawa coupling $g_{Y_1}$:
\be
Z_{Y_1}^{DR, \MSbar}= 1 + {\cal O}(g^4) .
\label{ZYDR}
\ee
Thus, we conclude that, at the quantum level, the Yukawa coupling renormalization in DR is not affected by one-loop corrections. Therefore, we anticipate that the corresponding renormalization on the lattice will be finite.

The renormalization matrix of the squark field, $Z_A$, is non-diagonal on the lattice, leading to mixing between the squark components due to the finite off-diagonal elements of $Z_A$. Furthermore, the use of Wilson discretization breaks the ${\cal \chi} \times {\cal R}$ symmetry, resulting in lattice bare Green's functions not being invariant under this symmetry at the quantum level. Consequently, one-loop spurious contributions will appear in the calculation of the lattice bare Green's functions, which must be eliminated by adding a mirror Yukawa counterterm to the action. 


By combining the difference between the one-loop $\MSbar$-renormalized Green's functions and the corresponding lattice bare Green's functions, the renormalization factors of fields and gauge coupling on the lattice, we obtain the following results: 
\bea
\label{zy1L}
{Z_{Y_1}}^{LR,\MSbar} &=& 1 + \frac{g^2}{16\,\pi^2}  \left(\frac{1.45833}{N_c} + 2.40768N_c + 0.520616 N_f \right),\\
{g_{Y_2}}^{LR,\MSbar} &=&\frac{g^3}{16\,\pi^2}\left(\frac{-0.040580}{N_c} + 0.45134 N_c \right).
\label{zy2L}
\eea
As anticipated by general renormalization theorems, the $\MSbar$-renormalization factors for gauge-invariant quantities are gauge-independent, which is also true in this case. Moreover, both the multiplicative renormalization factor, $Z_{Y_1}$, and the coefficient, $g_{Y_2}$, for the mirror Yukawa counterterm are finite, as predicted by the continuum calculation.

\section{Renormalization of the Quartic Couplings}
\label{couplingQ}

To renormalize the quartic couplings (four-squark interactions), we need to compute Green's functions involving four external squark fields. Two of these squarks must be in the fundamental representation, while the other two must be in the antifundamental representation. There are ten possible ways to select these squarks for $N_f > 1$. Some of these cases appear in the SQCD action and they will be used to obtain the renormalization factor of the quartic coupling, $Z_{\la_1}$. The other terms can appear as counterterms, in combinations which are invariant under all symmetries of the lattice SQCD action, including $\cal{C}$ and $\cal{P}$ symmetries. In this work, we examine $N_f$ fundamental multiplets, where the flavor symmetries permit the introduction of five "Fierz" operators that require fine-tuning. Ten combinations of quartic squark terms~\cite{Wellegehausen:2018opt}, are shown in Table~\ref{tb:nonsinglet2}.

\begin{table}[ht!]
\begin{center}
\scalebox{0.75}{
  \begin{tabular}{c|c|c}
\hline \hline
Operators & $\cal{C}$&$\cal{P}$ \\ [0.5ex] \hline \hline
    $\la_1 \left[(A_{+ \, f}^{\dagger} \, T^{\alpha} \, A_{+ \, f})(A_{+ \, f'}^{\dagger} \, T^{\alpha} \, A_{+ \, f'}) - 2 (A_{+ \, f}^{\dagger} \, T^{\alpha} \, A_{+ \, f}) (A_{-\,f'} \, T^{\alpha} \, A_{-\,f'}^{\dagger})  + (A_{-\,f} \, T^{\alpha} \, A_{-\,f}^{\dagger})(A_{-\,f'} \, T^{\alpha} \, A_{-\,f'}^{\dagger}) \right]/2 $&$+$& $+$   \\[0.75ex]\hline
    $\la_2 \left[(A_{+ \, f}^{\dagger}  A_{-\,f}^{\dagger})(A_{+\,f'}^{\dagger}  A_{-\,f'}^{\dagger})+ (A_{-\,f} A_{+ \, f})(A_{-\,f'} A_{+ \, f'})\right] $&$+$& $+$   \\[0.75ex]\hline
    $\la_3  (A_{+ \, f}^{\dagger}  A_{+ \, f})  (A_{-\,f'} A_{-\,f'}^{\dagger}) $&$+$ & $+$  \\[0.75ex]\hline
    $\la_4 (A_{+ \, f}^{\dagger}  A_{-\,f}^{\dagger}) (A_{-\,f'} A_{+ \, f'}) $&$+$& $+$   \\[0.75ex]\hline
    $\la_5 (A_{+ \, f}^{\dagger}  A_{-\,f}^{\dagger} + A_{-\,f} A_{+ \, f})(A_{+ \, f'}^{\dagger}  A_{+ \, f'}  + A_{-\,f'} A_{-\,f'}^{\dagger}) $&$+$ & $+$  \\[0.75ex]\hline
    $\la_1^F \left[(A_{+ \, f}^{\dagger} \, T^{\alpha} \, A_{+ \, f'})(A_{+ \, f'}^{\dagger} \, T^{\alpha} \, A_{+ \, f}) - 2 (A_{+ \, f}^{\dagger} \, T^{\alpha} \, A_{+ \, f'}) (A_{-\,f'} \, T^{\alpha} \, A_{-\,f}^{\dagger})  + (A_{-\,f} \, T^{\alpha} \, A_{-\,f'}^{\dagger})(A_{-\,f'} \, T^{\alpha} \, A_{-\,f}^{\dagger}) \right]/2 $&$+$& $+$   \\[0.75ex]\hline
    $\la_2^F \left[(A_{+ \, f}^{\dagger}  A_{-\,f'}^{\dagger})(A_{+\,f'}^{\dagger}  A_{-\,f}^{\dagger})+ (A_{-\,f} A_{+ \, f'})(A_{-\,f'} A_{+ \, f})\right] $&$+$& $+$   \\[0.75ex]\hline
    $\la_3^F  (A_{+ \, f}^{\dagger}  A_{+ \, f'})  (A_{-\,f'} A_{-\,f}^{\dagger}) $&$+$ & $+$  \\[0.75ex]\hline
    $\la_4^F (A_{+ \, f}^{\dagger}  A_{-\,f'}^{\dagger}) (A_{-\,f'} A_{+ \, f}) $&$+$& $+$   \\[0.75ex]\hline
    $\la_5^F (A_{+ \, f}^{\dagger}  A_{-\,f'}^{\dagger} + A_{-\,f} A_{+ \, f'})(A_{+ \, f'}^{\dagger}  A_{+ \, f}  + A_{-\,f'} A_{-\,f}^{\dagger}) $&$+$ & $+$  \\[0.75ex]\hline
\hline
\end{tabular}}
\caption{Dimension-4 operators which are gauge invariant and flavor singlets. All operators appearing in this table are eigenstates of charge conjugation, $\cal{C}$, and parity, $\cal{P}$, with eigenvalue 1. The flavor indices ($f$, $f'$) on the squark fields are explicitly displayed, while the color indices, which are the same within each parenthesis, are implicit; a summation is implied over all flavor and color indices. The parameters $\la_i$ and $\la_i^F$ denote the ten quartic couplings, while the superscript letter $F$ stands for Fierz.}
\label{tb:nonsinglet2}
\end{center}
\end{table}

\begin{figure}[ht!]
\centering
\includegraphics[scale=0.39]{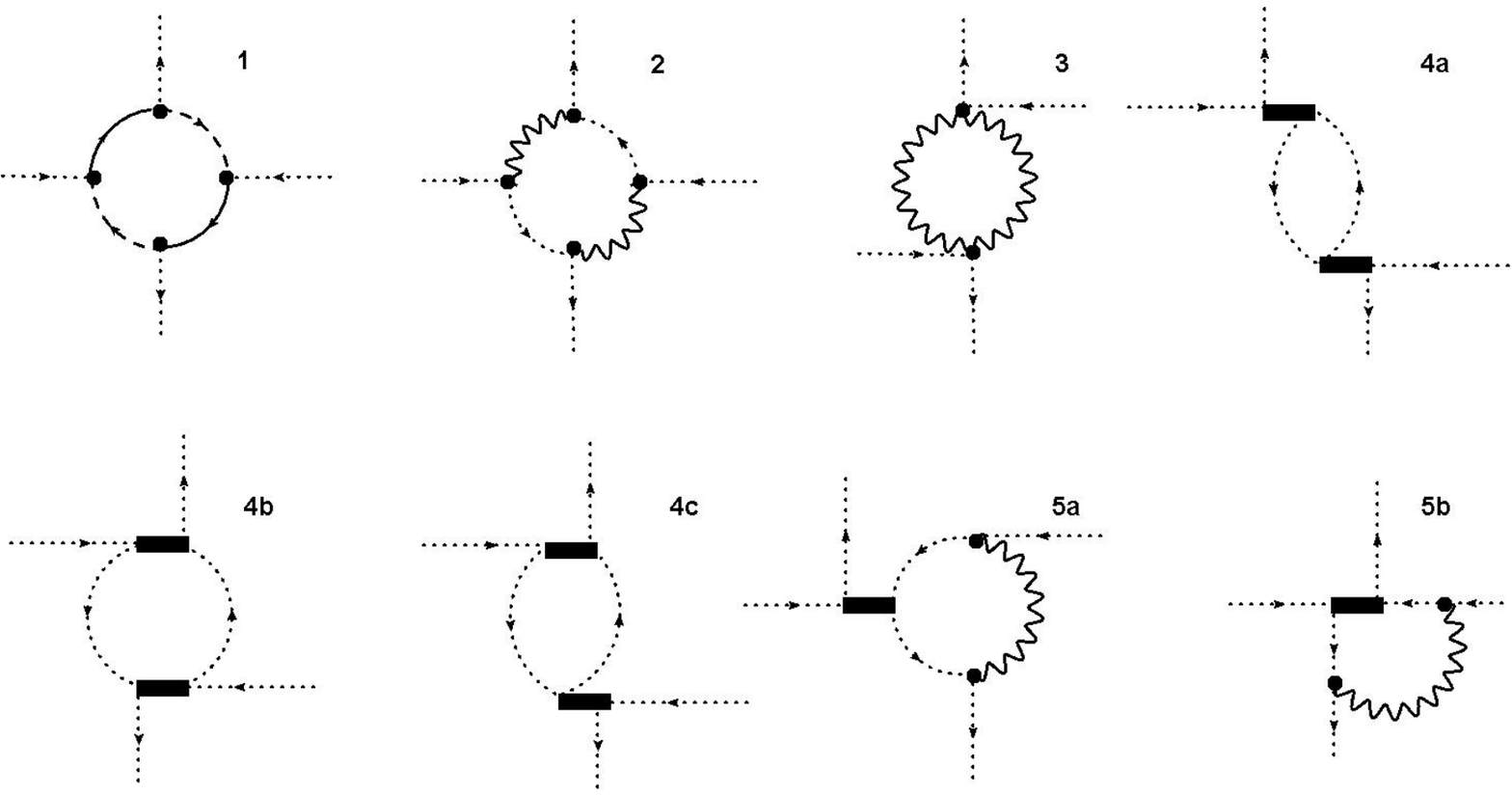} \\
\vspace{-2cm}
\hspace{0cm} \includegraphics[scale=0.32]{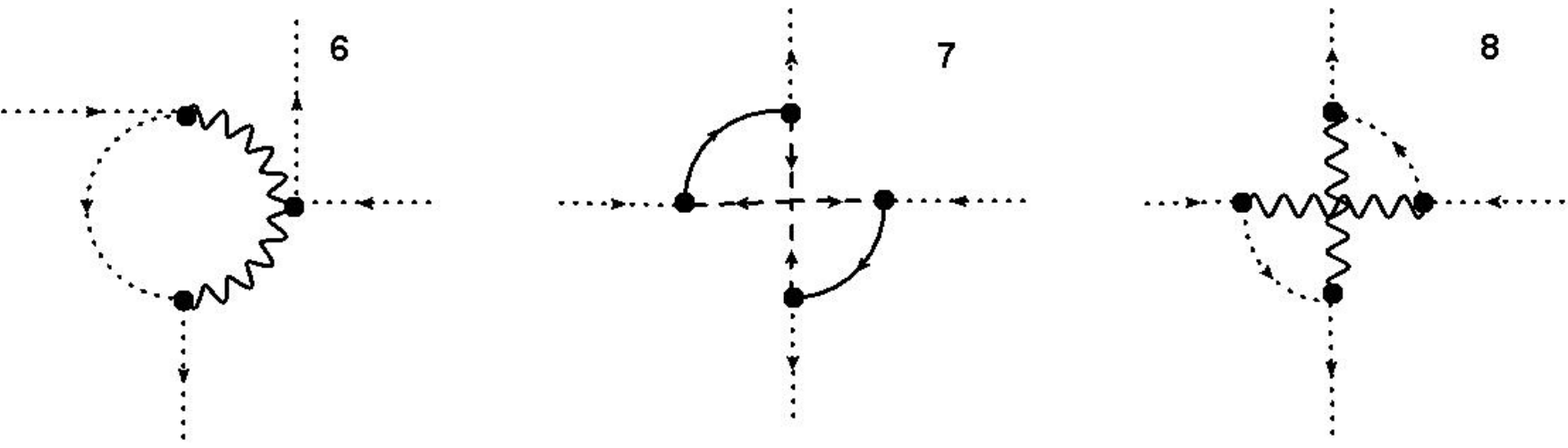}\\
\vspace{1cm}
\hspace{2cm}
\vspace{-3cm}\caption{One-loop 1PI Feynman diagrams leading to the fine-tuning of the quartic couplings.  A wavy (solid) line represents gluons (quarks). A dotted (dashed) line corresponds to squarks (gluinos). 
In the above diagrams the directions of the external line depend on the particular Green's function under study. An arrow entering (exiting) a vertex denotes a $\la, \psi, A_+, A_-^{\dagger}$ ($\bar \la, \bar \psi, A_+^{\dagger}, A_-$) field. The 4-squark vertex of the action has been denoted by a solid rectangle, in order to indicate the squark-antisquark color pairing; all remaining vertices are denoted by a solid circle. Squark lines could be further marked with a $+$($-$) sign, to denote an $A_+ \, (A_-)$ field.  All diagrams can have mirror variants. In diagrams 4 and 5, there are additional variants in which two external outgoing (or incoming) lines stem from a 4-squark vertex. }
\label{quarticFD1PI}
\end{figure}

\begin{figure}[ht!]
\centering
\vspace{-1cm}\includegraphics[scale=0.4]{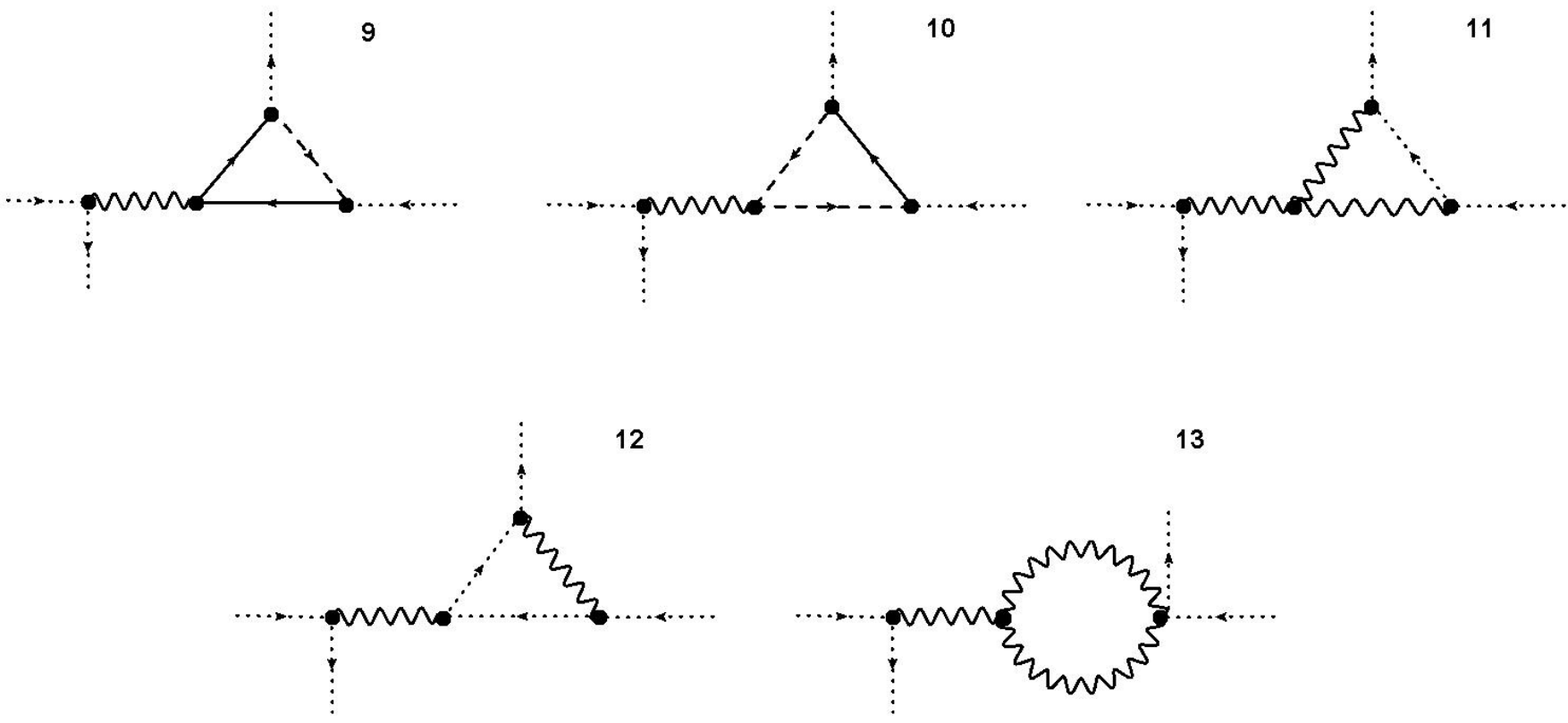}
\vspace{-2cm}
\newline
\hspace{-0.5cm}\includegraphics[scale=0.4]{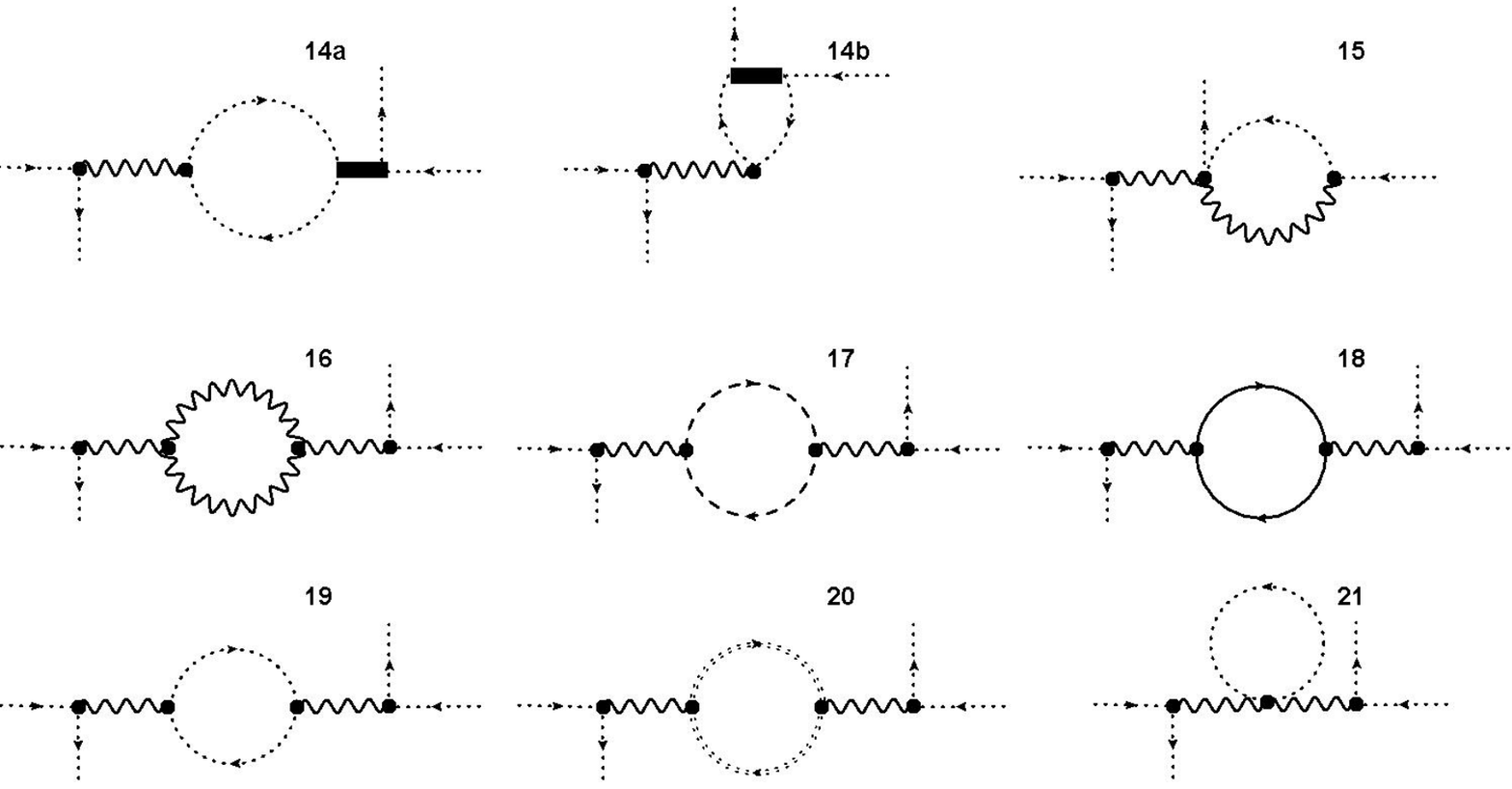}
\vspace{-1cm}\caption{One-loop 1PR Feynman diagrams leading to the fine-tuning of the quartic couplings.  Notation is identical to that of Figure \ref{quarticFD1PI}.  Note that the “double dashed” line is the ghost field. All diagrams can have mirror variants. 
}
\label{quarticFD1PR}
\end{figure}

\begin{figure}[ht!]
\centering
\vspace{-1cm}\includegraphics[scale=0.4]{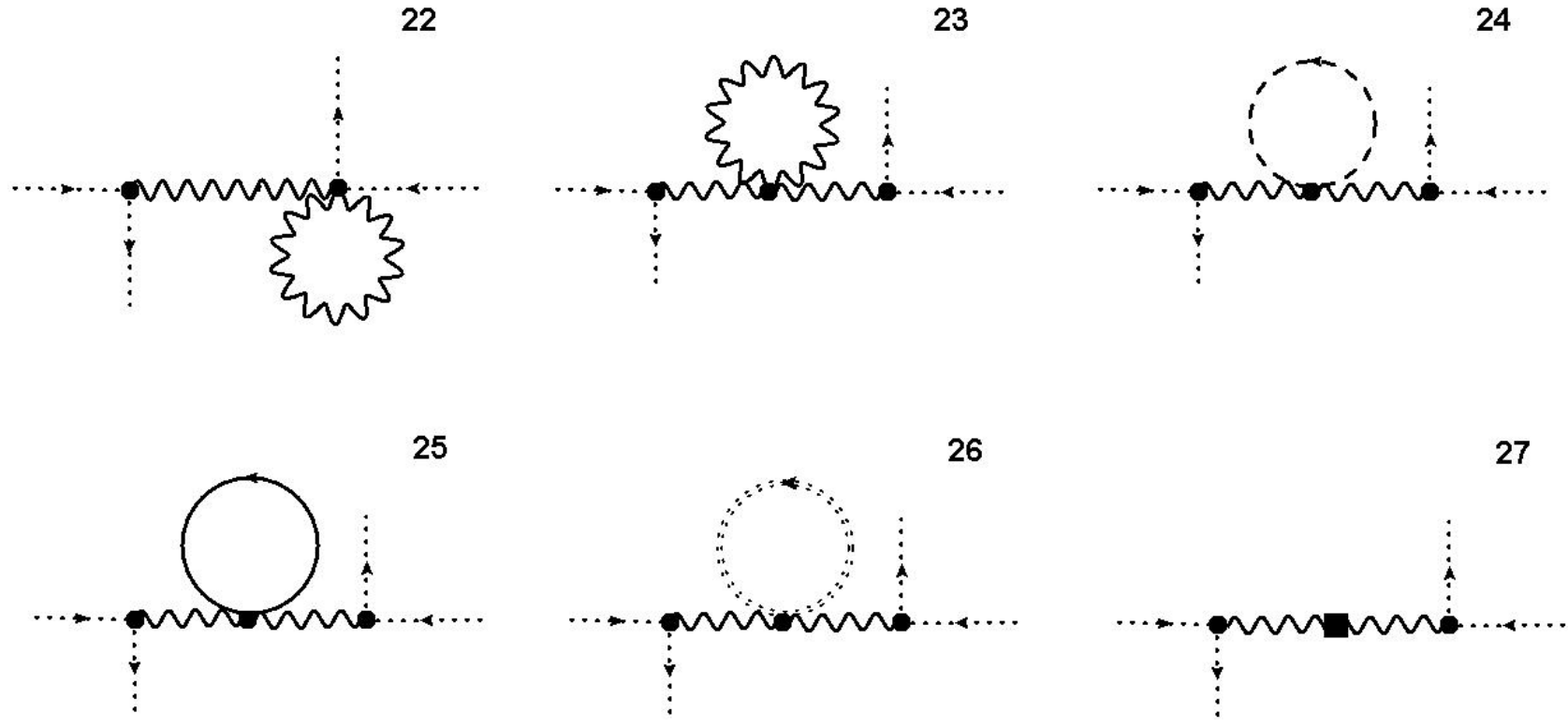}
\vspace{-1cm}\caption{Additional one-loop Feynman diagrams leading to the fine-tuning of the quartic couplings on the lattice. Notation is identical to that of Fig. \ref{quarticFD1PI}. Note that the “double dashed” line is the ghost field and the solid box in diagram 27 comes from the measure part of the lattice action.
}
\label{quarticFD}
\end{figure}

The tree-level values of the quartic couplings, $\la_i$ and $\la^F_i$, shown in Table~\ref{tb:nonsinglet2}, are:
\be
\la_1 =  g^2,  \, \, \, \la_2 = \la_3 = \la_4= \la_5 = \la^F_1 = \la^F_2 = \la^F_3 = \la^F_4= \la^F_5 = 0 \, .
\ee
These couplings receive quantum corrections, coming from the Feynman diagrams that are shown in Figs.~\ref{quarticFD1PI} and \ref{quarticFD1PR}. These one-loop diagrams enter the computation of the 4-pt amputated Green's functions for the quartic couplings. We compute, perturbatively, the relevant Green's functions using both dimensional and lattice regularization. Similar to the approach used for the Yukawa coupling~\cite{Costa:2023cqv}, for computational convenience, we can appropriately choose the external momenta, and after confirming the absence of superficial IR divergences, we simplify the calculations by setting the momenta of the two external squark fields in the fundamental representation to zero. There are also additional one-loop Feynman diagrams, which are illustrated in Fig.~\ref{quarticFD}, leading to the fine-tuning of the quartic couplings on the lattice.

Standard definitions of the renormalization factors of the fields, the gauge coupling constant, $Z_g$, the gauge parameter, $Z_{\alpha}$, and the quartic coupling, $Z_{\la_1}$, can be found in Ref.~\cite{Costa:2024tyz}. We impose renormalization conditions that eliminate divergences in the corresponding bare 4-pt amputated Green's functions. An example of such a renormalization condition, using DR, is also presented in Ref.~\cite{Costa:2024tyz}. By utilizing the renormalization conditions and the pole parts of the bare Green's functions, we derive the value for the renormalization factor of $\la_1^{DR, \MSbar}$:
\bea
{Z_{\la_1}}^{DR,\MSbar} &=& 1 + {\cal O}(g^4).
\label{ZlaDR}
\eea Eq.~(\ref{ZlaDR}) suggests that, at the quantum level, the quartic coupling renormalization in DR is not affected by one-loop corrections. It also implies that the corresponding renormalization on the lattice will be finite. While terms proportional to $\lambda_2$ - $\lambda_5$ and $\lambda^F_1$ - $\lambda^F_5$ do not appear in the $\MSbar$ renormalization within DR, a finite coefficient of these terms will arise on the lattice.

Shifting our focus to $LR$, it is worth mentioning that the lattice introduces mixing between squark components through the non-diagonal elements of the matrix $Z_A$, making the renormalization conditions more complex than those in DR. An example of a renormalization condition on the lattice can be found in Ref.~\cite{Costa:2024tyz}.

By combining the lattice expressions with the $\MSbar$-renormalized Green's functions computed in the continuum, we derive the following renormalization factor and counterterm coefficients:
\bea
{Z_{\la_1}}^{LR,\MSbar} &=& 1 + \frac{g^2}{16 \, \pi^2} \bigg( \frac{35.0365}{N_c} - 25.0526 \, N_c - 2.8298 \, N_f \bigg) \\
{\la_2}^{LR,\MSbar} &=&  \frac{ g^4}{16 \, \pi^2} \bigg[ -0.9488 \, \bigg( 1 + \frac{2}{N_c^2} \bigg) \bigg] \\
{\la_3}^{LR,\MSbar} &=& \frac{ g^4}{16 \, \pi^2}  \bigg(  \frac{28.427}{N_c^2 }\bigg) \\
{\la_4}^{LR,\MSbar} &=& \frac{g^4}{16 \, \pi^2} \bigg( 16.0381 + \frac{3.6493}{N_c^2} \bigg) \\
{\la_5}^{LR,\MSbar} &=& \frac{ g^4}{16 \, \pi^2} \bigg[ 0.4913 \, \bigg(1 + \frac{2}{N_c^2} \bigg) \bigg] \\
{\la^F_1}^{LR,\MSbar} &=& \frac{ g^4}{16 \, \pi^2} \bigg( 28.427  \bigg) \\
{\la^F_2}^{LR,\MSbar} &=& \frac{g^4}{16 \, \pi^2} \bigg[0.9488 \, \bigg( \frac{4}{N_c}  -  N_c \bigg) \bigg]\\
{\la^F_3}^{LR,\MSbar} &=& \frac{g^4}{16 \, \pi^2} \bigg(- \frac{21.5121}{N_c} + 1.8246 \, N_c \bigg) \\
{\la^F_4}^{LR,\MSbar} &=& \frac{g^4}{16 \, \pi^2} \bigg[ 14.2133 \, \bigg( -\frac{3}{N_c} +  N_c \bigg) \bigg] \\
{\la^F_5}^{LR,\MSbar} &=& \frac{g^4}{16 \, \pi^2} \bigg[0.4913 \, \bigg( -\frac{4}{N_c} +  N_c \bigg) \bigg].
\eea
It is important to note that the above factors are gauge-independent, as expected in the $\MSbar$ scheme. Furthermore, as predicted by the continuum computations, these couplings acquire only finite values.

We also provide results for the specific case where the number of flavors is $N_f=1$. In this context, we present the values for the five quartic terms, as no Fierz version exists with one flavor. Our results for these terms are as follows:
\bea
{Z_{\lambda_1}}^{LR,\overline{\text{MS}}}\Big{|}_{N_f=1} &=& 1 + \frac{g^2}{16 \, \pi^2} \bigg( -31.2567 + \frac{35.0366}{N_c} - 25.0526 \, N_c \bigg) \\
{\lambda_2}^{LR,\overline{\text{MS}}}\Big{|}_{N_f=1} &=&  \frac{g^4}{16 \, \pi^2} \bigg[-0.9488 \,\bigg( 1 + \frac{2}{N_c^2} - \frac{4}{N_c} + N_c \bigg)\ \bigg] \\
{\lambda_3}^{LR,\overline{\text{MS}}}\Big{|}_{N_f=1} &=& \frac{g^4}{16 \, \pi^2} \bigg(\frac{28.427}{N_c^2} - \frac{21.5121}{N_c} + 1.8246 \, Nc \bigg) \\
{\lambda_4}^{LR,\overline{\text{MS}}}\Big{|}_{N_f=1} &=& \frac{g^4}{16 \, \pi^2} \bigg[ 14.2133 \,  \bigg( 1 + N_c - \frac{3}{N_c}\bigg) + 1.8246 \, \bigg(1 + \frac{2}{N_c^2} \bigg) \bigg] \\
{\lambda_5}^{LR,\overline{\text{MS}}}\Big{|}_{N_f=1} &=& \frac{g^4}{16 \, \pi^2} \bigg[0.4913 \, \bigg( 1 + \frac{2}{N_c^2} - \frac{4}{N_c} +  N_c \bigg) \bigg].
\eea
These findings are crucial for understanding the behavior of the theory within this specific flavor sector.

\section{Summary -- Future Plans}
\label{summary}

In this work we calculate the renormalization factors for Yukawa and quartic couplings within the context of $\mathcal{N} = 1$ Supersymmetric QCD, in the Wess-Zumino gauge. To determine these factors, we compute the relevant 3-pt and 4-pt Green's functions using both dimensional and lattice regularizations. Our results indicate that the renormalization factor for the Yukawa and quartic couplings, as well as the coefficients of the mirror Yukawa counterterm and the quartic counterterms, are finite on the lattice. The findings of this project are crucial for setting up and calibrating lattice numerical simulations of SQCD. It is anticipated, in the coming years, that simulations of supersymmetric theories will become increasingly feasible and accurate.

A natural extension of this research involves perturbative calculations for all fine-tunings in SQCD and on the lattice, using chirally invariant actions. Specifically, the overlap action could be employed for gluino and quark fields to ensure proper chiral properties. While simulating overlap fermions presents a considerable challenge due to high CPU time requirements, it reduces the number of parameters that need fine-tuning, offering a significant advantage. However, determining the correct values for this minimal set of parameters still requires calculating a wide range of Green's functions.

\acknowledgments 
 The project (EXCELLENCE/0421/0025) is implemented under the programme of social cohesion ``THALIA 2021-2027'' co-funded by the European Union through the Research and Innovation Foundation of Cyprus (RIF). The results are generated within the FEDILA software (project: CONCEPT/0823/0052), which is also implemented under the same ``THALIA 2021-2027'' programme, co-funded by the European Union through RIF. M.C. also acknowledges partial support from the Cyprus University of Technology under the ``POST-DOCTORAL'' programme.

\end{document}